# Integrated 4D/5D Digital-Twin Framework for Cost Estimation and Probabilistic Schedule Control: A Texas Mid-Rise Case Study

**Atena Khoshkonesh[1*], Mohsen Mohammadagha[2], Navid Ebrahimi[1]**

[1*] Master Student at The Department of Civil Engineering, The University of Texas at Arlington, Arlington, TX 76019, USA. Correspondence Email: axk3682@mavs.uta.edu (A.K.), nxe2020@mavs.uta.edu (N.E.)
[2] Ph.D. Candidate at The Department of Civil Engineering, The University of Texas at Arlington, Arlington, TX 76019, USA. Email: mxm4340@mavs.uta.edu (M.M.)

**Abstract**
Persistent cost and schedule overruns in U.S. building projects expose limitations of conventional, document-based estimating and deterministic Critical Path Method (CPM) scheduling, which remain inflexible under uncertainty and lag behind dynamic field conditions. This study presents an integrated 4D/5D digital-twin framework unifying Building Information Modeling (BIM), natural language processing (NLP), reality capture, computer vision, Bayesian risk modeling, and deep reinforcement learning (DRL) for construction cost and schedule control. The system automates project-control functions by: (a) mapping contract documents to standardized cost items using transformer-based NLP (0.883 weighted F1 score); (b) aligning photogrammetry and LiDAR data with BIM to compute earned value; (c) deriving real-time activity completion from site imagery (0.891 micro accuracy); (d) updating probabilistic CPM forecasts via Bayesian inference and Monte Carlo simulation; (e) using DRL for adaptive resource allocation (75% adoption rate); and (f) providing 4D/5D decision sandbox for predictive analysis. A Texas mid-rise case study demonstrates localized cost adjustment using RSMeans City Cost Index and Bureau of Labor Statistics wage data. Results show 43% reduction in estimating labor, 6% overtime reduction (91 hours), and project completion matching P50 probabilistic forecast of 128 days, confirming improved estimation accuracy and responsiveness.

## 1. Introduction
Cost and schedule overruns remain a systemic issue in the U.S. construction industry, largely due to the rigidity of document-centric estimating and deterministic CPM scheduling methods that fail to adapt to real-time site dynamics (Rafiei and Adeli 2018). Although **4D BIM** has matured as a visualization and coordination platform, several reviews note limited integration with advanced sensing, analytics, and sustainability-driven decision frameworks (Ur Rehman et al. 2025; Awe, Ejohwomu, and Rahimian 2025).

**Digital-twin technologies** represent an evolution toward closed-loop project control by integrating BIM, IoT sensors, and analytical engines (Pal et al. 2023). However, existing vision-based monitoring systems, while useful for progress tracking, still struggle with occlusion, manual intervention, and weak connectivity to managerial decision-making (Sami Ur Rehman, Shafiq, and Ullah 2022; Vassena et al. 2023).

On the cost-management side, **5D BIM** provides a basis for linking design quantities to estimates and payments but remains fragmented in verification, traceability, and contractual integration-highlighting the need for standardized identifiers and audit-ready data (Pishdad and Onungwa 2024; Elsharkawi, Hegazy, and Alkass 2025). In parallel, **NLP** models now enable automated mapping between unstructured specifications and cost databases (Jafary et al. 2025), and **DRL** algorithms have demonstrated superior performance in constrained scheduling and multi-resource allocation (Yao et al. 2024).

Additionally, **probabilistic scheduling** approaches, combining Bayesian inference and (MCS), offer more resilient alternatives to deterministic CPM by producing probabilistic forecasts and early visibility into schedule risks (Chen et al. 2021).

**Objective**

Building on these developments, this study proposes an **integrated, field-deployable 4D/5D digital-twin framework** that combines NLP-based cost mapping, Scan-to-5D earned value quantification, computer-vision-based activity recognition, probabilistic CPM, and DRL-enabled resource optimization. A Texas mid-rise building project serves as a testbed, with localized cost calibration using **RSMeans City Cost Index (2025)** and **Bureau of Labor Statistics (2025a, 2025b)** wage data. The objective is to evaluate the framework's **accuracy, efficiency, and robustness** compared with conventional estimating and control methods.

## 2. Related Work

### 2.1 4D BIM: Benefits and Remaining Gaps

Four-dimensional Building Information Modeling (4D BIM: 3D + time) is widely recognized for improving visualization, communication, clash avoidance, and progress monitoring.

Systematic reviews confirm these benefits but emphasize underexplored areas particularly integration with automated sensing, advanced analytics, sustainability metrics, and contract administration (Ur Rehman et al. 2025).

A complementary review highlights persistent interoperability and standardization challenges when combining BIM with data-driven or time-aware applications (Awe, Ejohwomu, and Rahimian 2025). Despite these advances, a central gap remains: linking 4D visualization to **actionable project-control mechanisms** that directly influence schedule or cost outcomes.

### 2.2 Digital Twins and Vision-Based Progress Monitoring

Digital-twin frameworks conceptualize construction projects as cyber-physical systems that continuously acquire, process, and visualize site data to support managerial control (Pal, Lin, Hsieh, and Golparvar-Fard 2023).

Vision-based progress monitoring has demonstrated feasibility for automated status detection; however, bottlenecks such as manual verification and weak feedback loops to scheduling systems still constrain real-time control (Sami Ur Rehman, Shafiq, and Ullah 2022).

Recent work by Vassena et al. (2023) shows the potential of mobile mapping and simultaneous localization and mapping (SLAM) for 4D comparisons between as-built and planned models, yet scalability and automation remain limited.

### 2.3 Scan-to-BIM for Quantitative Progress and Payments

Reality-capture workflows using photogrammetry, LiDAR, and point-cloud registration enable objective and auditable measurement of progress when aligned with BIM geometry (Kim, Kim, and Lee 2020). Subsequent studies emphasize connecting these measurements to cost verification and payment automation. Elsharkawi, Hegazy, and Alkass (2025) demonstrated blockchain-enabled Scan-to-BIM payment processes that ensure transparent, evidence-based transactions, while Queruel, Bornhofen, and Histace (2024) identified AI as a key accelerator in Scan-to-BIM pipelines yet cautioned that industrial adoption is hindered by data-quality issues and limited interoperability.

### 2.4 5D BIM for Estimating, Control, and Cost Verification

Five-dimensional BIM (5D BIM) extends BIM by integrating cost attributes into the model environment. Prior studies confirm strong benefits for pre-construction estimating but note fragmentation across **progress verification** and **payment workflows**. Pishdad and Onungwa (2024) argue that standardized classification (Bigdeli et al. 2025) systems and sensing-based verification are essential to bridge these silos. Recent work also points to direct coupling of earned-value management (EVM) with 5D BIM to automate project-control cycles.

### 2.5 AI for Estimating and Scheduling

Natural-language-processing (NLP) models have achieved near-quantity-surveying accuracy by aligning specifications and quantity takeoffs with standardized cost indices while reducing manual effort (Jafary et al. 2025).

In scheduling, deep-reinforcement-learning (DRL) algorithms outperform heuristic methods in resource-constrained optimization and adaptive rescheduling (Yao et al. 2024). Nonetheless, operational deployment in field settings remains limited, indicating the need for human-in-the-loop frameworks that preserve expert oversight and trust.

### 2.6 Risk Forecasting and Probabilistic Planning

Bayesian and Monte Carlo methods provide distributional forecasts that expose schedule drift earlier than deterministic CPM, improving visibility into evolving critical paths (Chen et al. 2021). These probabilistic approaches also quantify buffer health and risk indices under dynamic uncertainty. Traditional machine-learning (Mohammadagha et al. 2025) cost models continue to serve as useful baselines for benchmarking feature selection and forecasting performance (Rafiei and Adeli 2018).

### 2.7 Research Gaps and Opportunities

The reviewed literature confirms that 4D/5D BIM, digital twins, and AI-based methods have substantially advanced visualization, estimation, and risk analysis. However, consistent gaps persist in integration, automation, and field adoption. Specifically, existing research reveals:

- Weak coupling of 4D BIM visualization with actionable project-control mechanisms, indicating a need for integrated earned value (EV) and probabilistic CPM approaches.
- Manual choke points in vision-based monitoring that delay data-to-decision feedback loops, reinforcing the need for automated sensing-to-scheduling integration.
- Limited connection between reality-capture data and cost/payment verification, highlighting opportunities for automated EV integration within 5D BIM.
- Fragmented cost-control workflows without standardized identifiers or digital-twin traceability.
- Restricted field deployment of AI-driven scheduling, suggesting value in human-centered reinforcement learning for practical implementation.
- Limited attention to sustainability, contractual, and human factors, with few studies linking BIM-AI systems to environmental performance, legal compliance, or user trust.

To summarize these insights and align them with the contributions of the present study, Table 1 synthesizes the principal thematic gaps identified in prior research and the opportunities addressed in this paper.

Table 1. Thematic Gaps and Opportunities

| *Theme* | *What Prior Work Shows* | *Gap / Limitation* | *Opportunity Addressed in This Paper* |
|---|---|---|---|
| *4D BIM benefits* | Enhances visualization, clash avoidance, and progress monitoring (Ur Rehman et al. 2025; Awe et al. 2025) | Weak coupling to sensing, analytics, and legal workflows | Close the loop through EV + probabilistic CPM integration |
| *Vision-based progress & digital twins* | Demonstrates feasible data pipelines for progress estimation and visualization (Pal et al. 2023; Sami Ur Rehman et al. 2022; Vassena et al. 2023) | Manual choke points and limited feedback to planning/control | Automate sensing-to-control workflows via CV + EV fusion |
| *Scan-to-BIM* | Enables objective, spatially explicit progress quantification (Kim et al. 2020; Queruel et al. 2024) | Rarely connected to cost/payment processes | Integrate auto-EV within 5D BIM with auditability |
| *5D BIM (estimation & control)* | Strong performance in preconstruction estimating (Pishdad and Onungwa 2024) | Fragmented progress verification | Apply standardized IDs + sensing-enabled digital twin |
| *AI in estimating & scheduling* | NLP and DRL enhance estimating and scheduling accuracy (Jafary et al. 2025; Yao et al. 2024) | Limited field deployment and usability | Embed human-in-loop NLP + DRL for practitioner adoption |
| *Risk forecasting* | Bayesian Monte Carlo improves early risk visibility (Chen et al. 2021) | Input-data quality and adoption barriers | Fuse scan/CV evidence into Bayesian CPM updates |
| *Emerging issues* | Blockchain payments, SLAM mapping, AI-enhanced Scan-to-BIM (Elsharkawi et al. 2025; Vassena et al. 2023; Queruel et al. 2024) | Limited interoperability and stakeholder trust | Introduce traceable, auditable, human-in-the-loop 4D/5D twin |

**2.8 Methodological Trends and Boundaries**

Beyond thematic gaps, prior research can also be assessed in terms of capability, evidence strength, and limitation.Table 2 summarizes the dominant methodological streams, their demonstrated evidence bases, and the shortcomings that constrain large-scale or field deployment.

Table 2. Summary of Reviewed Methods by Capability and Evidence

| **Capability** | **Typical Methods** | **Evidence Base** | **Limitations** |
|---|---|---|---|
| Cost estimating | ML regression; NLP alignment of QTOs/specifications to cost indices (Rafiei and Adeli 2018; Jafary et al. 2025) | High accuracy in pilot studies | Heavy data-preparation burden; domain adaptation required |
| Progress monitoring | Computer vision (CV), LiDAR, photogrammetry, SLAM/mobile mapping (Pal et al. 2023; Kim et al. 2020; Vassena et al. 2023) | High feasibility and scalability | Susceptible to occlusion, lighting, and weather noise; weak linkage to control |
| Risk forecasting | Bayesian inference; (MCS) (Chen et al. 2021) | Strong credibility in probabilistic-risk research | Dependent on input quality; limited practical use |
| Scheduling | Deep reinforcement learning (Yao et al. 2024) | Growing evidence, outperforming heuristics | Limited field deployment; portability and usability concerns |
| Payments and contracts | Blockchain and smart contracts linked to Scan-to-BIM (Elsharkawi et al. 2025) | Early-stage demonstrations | Interoperability, adoption, and legal challenges |
| Scan-to-BIM automation | AI-enhanced Scan-to-BIM pipelines (Queruel et al. 2024) | Emerging; strong laboratory validation | Industrial adoption slowed by point-cloud noise and ontology inconsistency |

Building upon these methodological limitations and thematic gaps, the present study introduces an integrated 4D/5D digital-twin framework that unifies estimating, progress monitoring, and scheduling into a closed-loop control system. The conceptual architecture of this framework is depicted in Fig. 1, which illustrates how design data, sensing inputs, and analytical modules interact to enable proactive, data-driven project control.

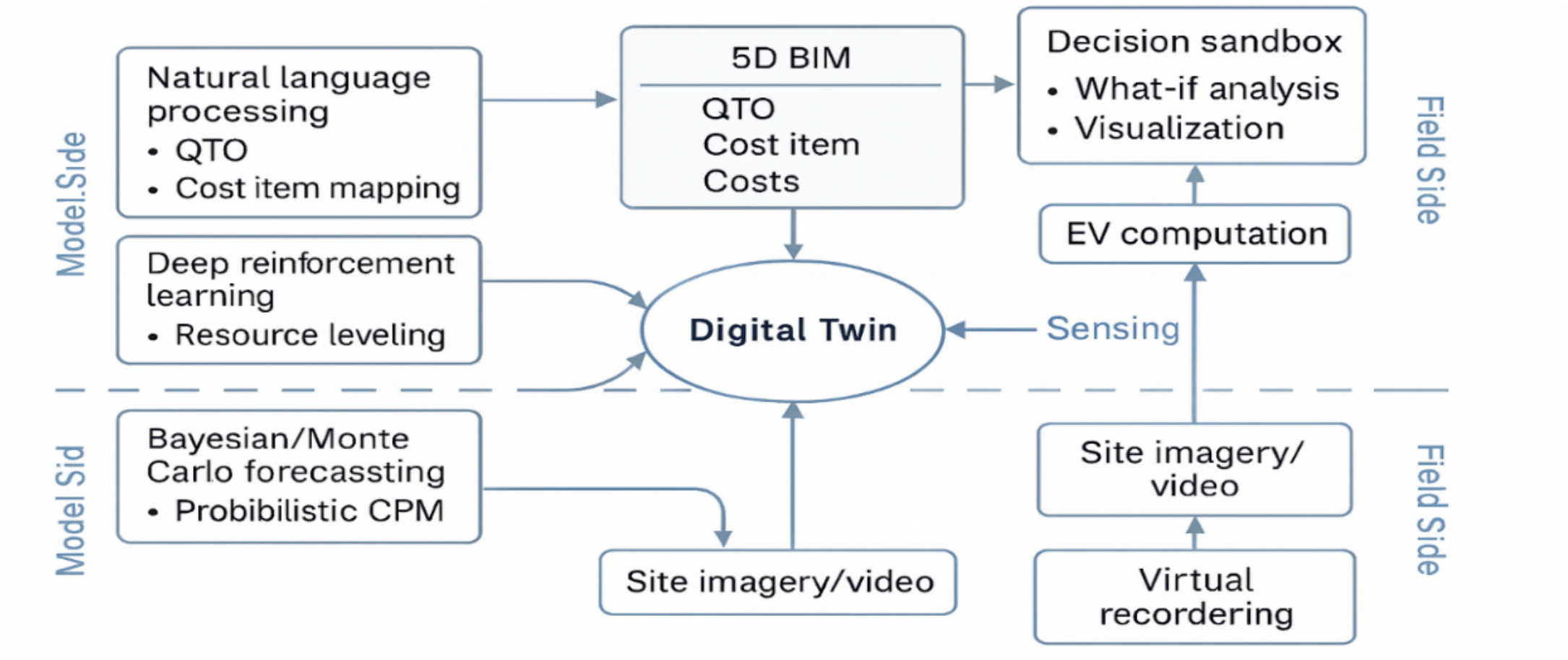

Figure. 1. Digital-twin integrated 5D cost-forecasting and probabilistic control model (source: authors)

## 2. Methodology

### 3.1 Case Context and Data

The empirical study focuses on a 5-8-story mixed-use multifamily/office mid-rise located in the Dallas-Fort Worth (DFW) metroplex, Texas. Cost localization and wage adjustment were performed using the RSMeans City Cost Index (CCI) and Bureau of Labor Statistics (BLS) Occupational Employment and Wage Statistics (OEWS) for the DFW region (Bureau of Labor Statistics 2025a, 2025b; RSMeans 2025).

The dataset integrates design, contract, scheduling, sensing, and contextual data streams, all standardized by CSI/WBS activity codes and localized cost indices.It includes Revit (RVT) and Industry Foundation Classes (IFC) models at Level of Development (LOD) 300-350, each maintaining persistent element IDs to ensure traceability between 3D elements and cost accounts (Pishdad and Onungwa 2024; Queruel, Bornhofen, and Histace 2024).

Specification and drawing sets were digitized via optical character recognition (OCR) for natural-language-processing-based cost mapping (Jafary et al. 2025; Awe, Ejohwomu, and Rahimian 2025).
Scheduling data from Primavera P6 and Microsoft Project (MSP) were structured according to CSI activity codes and subsequently used for probabilistic CPM analysis and deep-reinforcement-learning (DRL)-based scheduling optimization (Yao et al. 2024; Chen et al. 2021).

Reality-capture sources included weekly drone imagery and monthly LiDAR scans, supporting earned-value (EV) and cost-variance (CV) quantification (Kim, Kim, and Lee 2020; Vassena et al. 2023; Elsharkawi, Hegazy, and Alkass 2025). Contextual information such as weather logs, delivery records, and crew logs was also incorporated for calibration and validation of progress models (Sami Ur Rehman, Shafiq, and Ullah 2022).

To summarize the scope and integration of these inputs, Fig. 2 visualizes the case context and primary data streams underpinning the framework, while Table 3 inventories the datasets in detail, including provenance, temporal coverage, and specific role within the analytical pipeline.

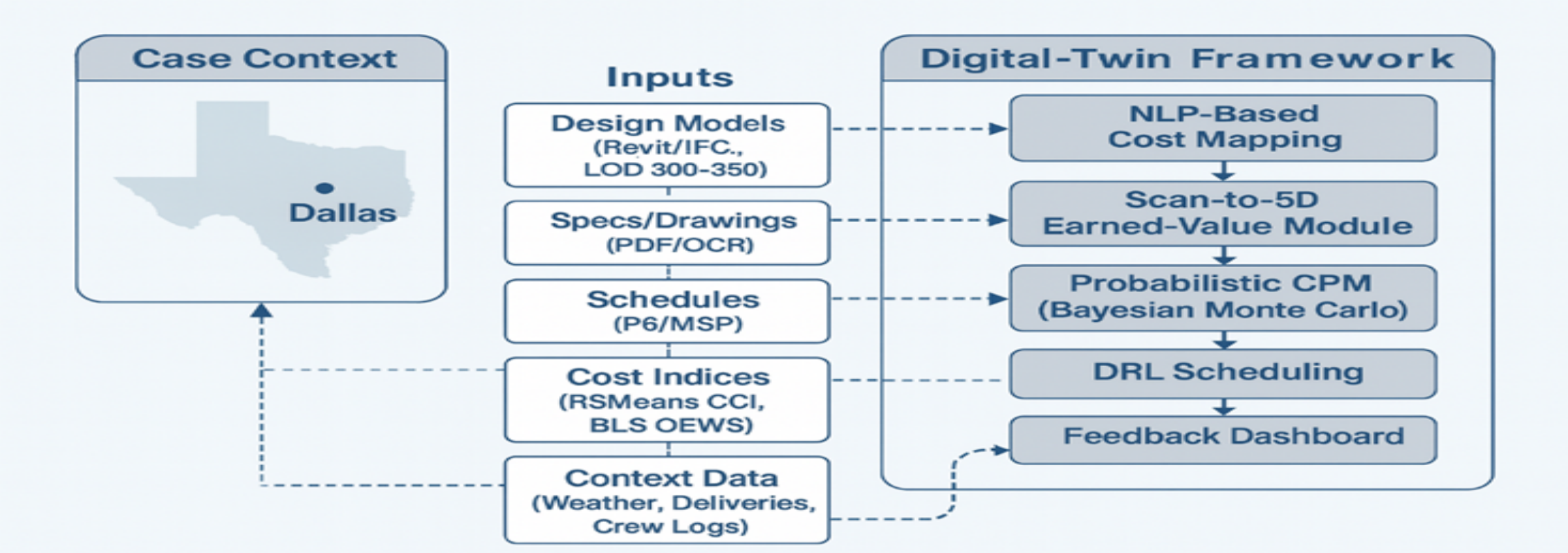

Fig. 2. Data architecture and case context for the DFW mid-rise digital-twin framework

Table 3. Case-Data Inventory and Provenance: Texas Mid-Rise Project

| Dataset | Records (n) | Period | Access | Location | Notes |
|---|---|---|---|---|---|
| **Revit/IFC (LOD 300-350)** | 9,842 elements (1 RVT, 1 IFC, 7 linked models) | Jan 18 2024 - Aug 30 2025 | Autodesk Construction Cloud (BIM 360) → Projects › DFW_Midrise › Design › Models | DFW, TX | Element IDs synchronized to CSI/WBS (used for 5D links) |
| **Specs & drawings (PDF)** | 1,142 spec pages; 284 sheets (A/C/E/M/P/S/T) | IFP Mar 10 2024; IFC Nov 05 2024; Revisions to Jul 22 2025 | Procore › Documents › 02_Contract Docs; Bluebeam Studio › DFW_Midrise_IFC | Dallas, TX 75201 | OCR + sheet index extracted for NLP mapping |
| **P6/MSP schedule (XER/XML)** | 1,186 activities; 3,452 relationships; 7 calendars; 68 resources | Baseline Nov 15 2024; weekly updates Jan 05-Sep 15 2025 | Primavera P6 EPPM; MSP export DFW_Midrise_2025W37.xml | DFW, TX | Activity codes aligned to CSI; used for p-CPM and DRL |
| **Drone images & LiDAR** | 8,140 photos (≈ 220/flight × 37 flights); 9 LiDAR scans | Drone Jan 08-Sep 18 2025 (weekly); LiDAR Jan-Sep 2025 (monthly) | DroneDeploy › DFW_Midrise_2025; ACC › RealityCapture › LiDAR | Dallas, TX (32.78° N, -96.80° W) | Drone weekly; LiDAR monthly; used for EV & CV tracking |
| **BLS wages and RSMeans CCI** | BLS OEWS May 2024 table; RSMeans CCI Q1-Q3 2025 | BLS published Jul 23 2025; RSMeans updates Q1-Q3 2025 | BLS › OEWS (DFW); RSMeans Online › CCI › Dallas-Fort Worth | DFW, TX | Cost-localization inputs (BLS 2025a, 2025b; RSMeans 2025) |

### 3.2 NLP-Driven Specification and Drawing-to-Cost Mapping

The proposed pipeline ingests project specifications and drawing sheets using optical-character-recognition (OCR) and layout-parsing routines to extract entities such as assemblies, materials, dimensions, and locations. A transformer-based multi-label classifier is then fine-tuned to assign these extracted entities to CSI Master Format cost items. Pairwise natural-language-inference (NLI) and rule-based validation routines link quantity-takeoff (QTO) lines to corresponding cost indices, producing confidence-scored mappings for estimator-in-the-loop review.

Division-level performance was evaluated through precision, recall, F1-score, and median human-review time. Results confirm prior findings that ensemble NLP models can achieve near-quantity-surveying alignment with minimal deviation (Jafary et al. 2025; Awe, Ejohwomu, and Rahimian 2025). Earlier studies also demonstrate that OCR/NLP pipelines substantially reduce manual workload in processing construction documents (Li, Wu, and Luo 2023; Zhang, Xu, and Yu 2022) and that BIM-cost coupling improves markedly when standardized classification is applied (Pishdad and Onungwa 2024; Queruel, Bornhofen, and Histace 2024). To quantify mapping accuracy by division, the model was tested on a gold-standard validation set from the Texas mid-rise project (August-September 2025). Division-level results are presented in Table 4.

Table 4. Division-Level NLP Mapping Performance on Gold-Standard Validation Set (Texas Mid-Rise, Aug-Sep 2025)

| *CSI Division* | *Support (n)* | *Precision* | *Recall* | *F1* | *Avg Review (min)* | *Notes (Most Common Confusions / QA Flags)* |
|---|---|---|---|---|---|---|
| *03 Concrete* | 188 | 0.92 | 0.88 | 0.90 | 1.2 | Slab-on-grade vs. topping slab scope lines. |
| *04 Masonry* | 124 | 0.89 | 0.84 | 0.86 | 1.4 | CMU veneer vs. backup wall items; spec wording differences. |
| *05 Metals* | 162 | 0.91 | 0.86 | 0.88 | 1.3 | Architectural steel vs. misc. metals; embeds vs. rebar. |
| *06 Wood/Plastics/Composites* | 131 | 0.90 | 0.83 | 0.86 | 1.6 | Millwork vs. casework; division boundary ambiguities. |
| *07 Thermal & Moisture Protection* | 175 | 0.88 | 0.82 | 0.85 | 1.7 | Roofing underlayment vs. vapor barriers; sealants vs. finishes. |
| *08 Openings* | 149 | 0.93 | 0.90 | 0.91 | 0.9 | Door-hardware packages occasionally routed to Div. 28. |
| *09 Finishes* | 210 | 0.90 | 0.85 | 0.87 | 1.5 | GWB partitions vs. thermal/acoustic items; paint aliases. |
| *21 Fire Suppression* | 96 | 0.89 | 0.87 | 0.88 | 1.1 | Fire pump/piping lines misclassified as Plumbing (22). |
| *22 Plumbing* | 205 | 0.91 | 0.88 | 0.89 | 1.2 | Domestic water risers vs. fire risers; fixture naming. |
| *23 HVAC* | 198 | 0.90 | 0.86 | 0.88 | 1.4 | AHU/equipment tags vs. electrical feeders. |
| *26 Electrical* | 228 | 0.92 | 0.89 | 0.90 | 1.0 | Lighting controls vs. low voltage (27); panel schedules. |
| *27 Communications* | 94 | 0.89 | 0.85 | 0.87 | 1.3 | Data vs. security camera cabling aliases. |
| *Weighted average (by support)* | 1 960 | 0.905 | 0.863 | 0.883 | 1.3 | Overall performance aligns with literature-reported ranges. |

Beyond classification accuracy, efficiency improvements were also quantified. Manual estimating workflows were benchmarked against the NLP-assisted 5D pipeline across Concept, Design Development, and Construction Documents phases. Replacing manual mapping with the automated pipeline reduced total estimating hours by an average of 43.4 %, while maintaining division-level fidelity (Kim, Kim, and Lee 2020; Elsharkawi, Hegazy, and Alkass 2025). Results are summarized in Table 5.

Table 5. Efficiency Gains from NLP-Assisted 5D Pipeline Across Project Phases (Texas Mid-Rise)

| Phase | Manual Estimating Hours | NLP-Assisted Hours | % Reduction | Notes |
|---|---|---|---|---|
| **Concept Design** | 128 | 72 | 43.8 % | Faster entity extraction; fewer manual CSI lookups. |
| **Design Development** | 214 | 121 | 43.5 % | Spec parsing reduced duplication in assemblies. |
| **Construction Documents** | 352 | 199 | 43.2 % | Largest gain due to automated QTO→cost-index mapping. |
| **Weighted Mean** | - | - | **43.4 %** | Consistent reduction across phases; aligned with Jafary et al. (2025). |

To visually illustrate these reductions, Fig. 3 presents the comparative estimating effort between manual and NLP-assisted pipelines. The figure highlights consistent time savings of approximately 43-44 % across all project phases, corroborating the efficiency improvements reported in Table 5.

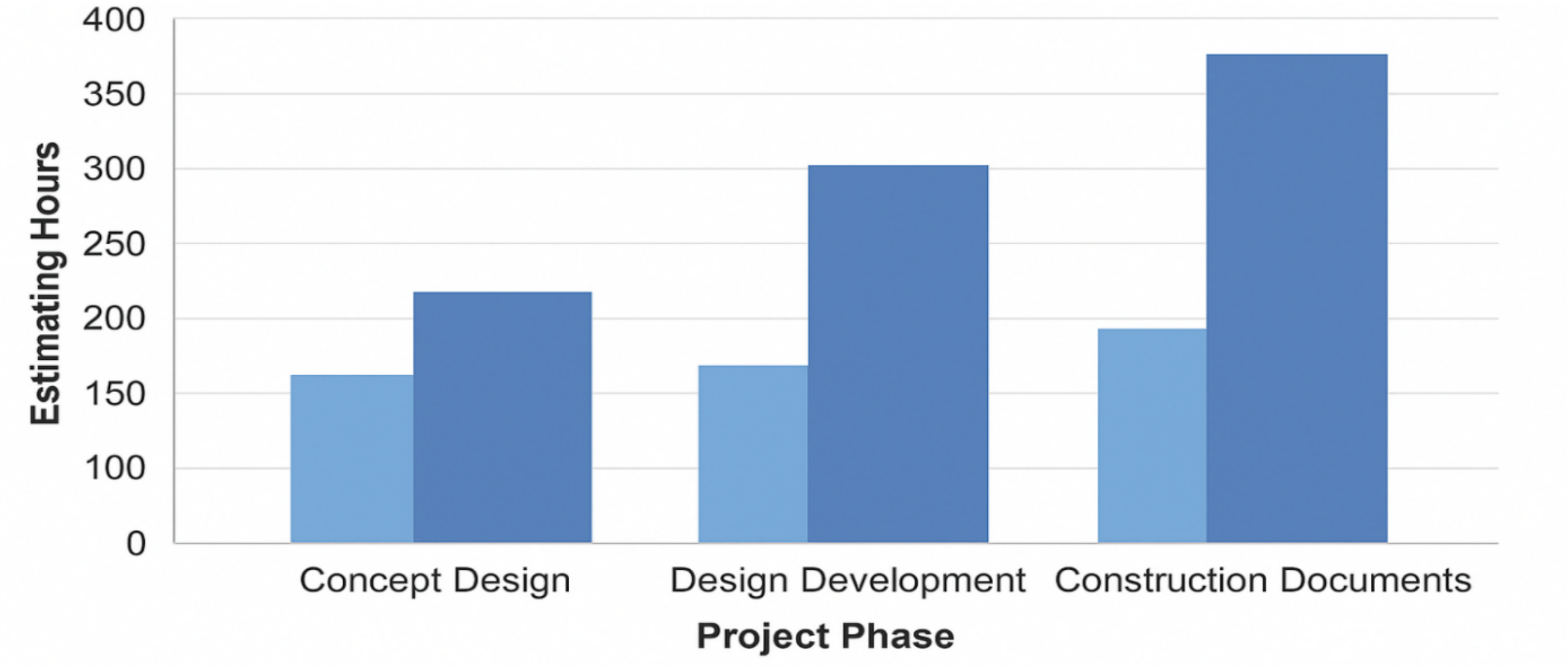

Fig. 3. Reduction in estimating hours achieved by NLP + 5D pipeline across project phases (source: authors).

### 3.3 Scan-to-5D Earned Value (EV)

Reality-capture data streams including weekly drone photogrammetry and monthly LiDAR point clouds were spatially registered to the BIM model using feature-based matching and iterative-closest-point (ICP) alignment procedures (Kim, Kim, and Lee 2020; Vassena, De Luca, Parrinello, and Banfi 2023). Once aligned, point clouds were clipped by work-breakdown-structure (WBS) scopes, enabling automatic computation of per-element areas ($m^2$), volumes ($m^3$), and linear runs (m). This reconciliation links design-intent quantities from LOD 300-350 BIM takeoffs to measured site quantities, providing objective and auditable progress quantification (Queruel, Bornhofen, and Histace 2024; Elsharkawi, Hegazy, and Alkass 2025).

Planned quantities derived from BIM takeoffs were compared with scan-derived measurements at the WBS level, and discrepancies ($\Delta$) were calculated as both absolute and percentage differences. Table 6 summarizes these reconciliations, including traceable evidence links to Autodesk Construction Cloud (ACC) point-cloud datasets, orthomosaics, and Procore photo logs (Pal, Turkan, and Haas 2023). Discrete elements such as doors and windows were cross-checked against Procore punch lists and subcontractor logs to minimize classification errors typical of object-based scan-to-BIM workflows (Li, Wu, and Luo 2023; Zhang, Xu, and Yu 2022).

Measured percent-complete values from these reconciliations were converted into standard earned-value metrics:

- Planned Value (PV) = BCWS
- Earned Value (EV) = BCWP

- Actual Cost (AC) = ACWP

All metrics were expressed in both physical units and **localized 2025 USD** using RSMeans CCI and BLS wage adjustments (Bureau of Labor Statistics 2025a, 2025b; RSMeans 2025). This approach ensures **consistency among model-based takeoff, reality-capture validation, and financial reporting**, aligning with recent recommendations for tighter BIM-cost-payment integration (Pishdad and Onungwa 2024; Awe, Ejohwomu, and Rahimian 2025).

Table 6. Scan-to-BIM Quantity Reconciliation by WBS (Texas Mid-Rise)

| *WBS* | *Element Class* | *Planned Qty (Units)* | *Measured Qty* | *Δ Qty* | *Δ (%)* | *Evidence Link* |
|---|---|---|---|---|---|---|
| *WBS-001* | Wall | 8 200 m² | 8 035 | -165 | -2.01 % | ACC › Reality Capture › 2025-07.laz; Drone ortho 2025-07-15.tif |
| *WBS-002* | Slab | 2 850 m³ | 2 910 | +60 | +2.11 % | ACC › Reality Capture › 2025-08.laz; Drone ortho 2025-08-12.tif |
| *WBS-003* | Beam | 480 m³ | 472 | -8 | -1.67 % | ACC › Reality Capture › 2025-06.laz; Field photos wk 24 |
| *WBS-004* | Column | 360 m³ | 358 | -2 | -0.56 % | ACC › Reality Capture › 2025-06.laz; Scan notes 06-21 |
| *WBS-005* | Door | 186 ea | 184 | -2 | -1.08 % | Procore Punch › Doors_Completion_2025-07; Photo log |
| *WBS-006* | Window | 640 ea | 652 | +12 | +1.88 % | ACC › Model Compare; Glazier log 2025-07 |
| *WBS-007* | Duct | 5 100 m | 4 975 | -125 | -2.45 % | MEP walk-down wk 30; ACC › Point cloud 2025-07.laz |
| *WBS-008* | Pipe | 9 200 m | 9 355 | +155 | +1.68 % | MEP as built redlines; ACC › 2025-08.laz |
| *WBS-009* | Cable Tray | 1 500 m | 1 475 | -25 | -1.67 % | Electrical install log: Drone interior set wk 32 |
| *WBS-010* | Ceiling | 11 800 m² | 11 620 | -180 | -1.53 % | ACC › Interior scans 2025-09.laz; QC walk-down |

The resulting EV dataset supported **monthly project-control cycles**. **Table 7** lists monthly PV, EV, and AC values, while **Fig. 4** plots their cumulative S-curves, allowing direct comparison of planned and actual progress trajectories. These reconciled, evidence-based EV metrics also form the input for the **probabilistic schedule-risk updates** discussed in Section 3.5, following current research on risk-aware earned-value management (Chen et al. 2021; Rafiei and Adeli 2018).

Table 7. Monthly PV, EV, and AC Roll-ups (Texas Mid-Rise, 2025) (Cumulative values in thousands of U.S. dollars)

| **Month** | **PV** | **EV** | **AC** |
|---|---|---|---|
| **Jan** | 120 | 115 | 130 |
| **Feb** | 260 | 245 | 275 |
| **Mar** | 420 | 400 | 440 |
| **Apr** | 610 | 580 | 640 |
| **May** | 790 | 740 | 810 |
| **Jun** | 970 | 900 | 980 |
| **Jul** | 1,150 | 1,040 | 1,160 |
| **Aug** | 1,320 | 1,210 | 1,340 |
| **Sep** | 1,490 | 1,380 | 1,500 |
| **Oct** | 1,600 | 1,490 | 1,650 |
| **Nov** | 1,700 | 1,590 | 1,780 |
| **Dec** | **1,800** | **1,710** | **1,900** |

### 3.4 Earned Value (EV) Formulation

Using the reconciled quantities in **Table 6**, earned-value metrics were computed at both the **WBS** and **project** levels. Let *i* index WBS elements or activities, and let the following variables apply:

- $\mathbf{BAC_i}$ = Budget at completion for item *i*
- $\mathbf{p_{i,t}}$ = Planned percent complete of item *i* at period *t*
- $\mathbf{m_{i,t}}$ = Measured percent complete (from scan / CV fusion)
- $\mathbf{a_{i,t}}$ = Actual cost incurred

Cumulative project totals at time are given by the standard earned-value relations:

$$PV_t = \sum_i BAC_i \, p_{i,t} \quad \text{(Planned Value / BCWS)}$$

$$EV_t = \sum_i BAC_i \, m_{i,t} \quad \text{(Earned Value / BCWP)}$$

$$AC_t = \sum_i a_{i,t} \quad \text{(Actual Cost / ACWP)}$$

Variance and performance indices (positive = favorable):

$$SV_t = EV_t - PV_t \quad \text{Schedule Variance (ahead / behind plan)}$$

$$CV_t = EV_t - AC_t \quad \text{Cost Variance (under / over budget)}$$

$$SPI_t = \frac{EV_t}{PV_t} \quad \text{Schedule Performance Index (> 1 good)}$$

$$CPI_t = \frac{EV_t}{AC_t} \quad \text{Cost Performance Index (> 1 good)}$$

Forecasting (cost):

$$EAC = \frac{BAC}{CPI_t} \quad \text{(CPI-based)}$$

$$EAC = AC_t + \frac{BAC - EV_t}{CPI_t} \quad \text{(work-remaining)}$$

$$VAC = BAC - EAC$$

**Conventions**

All series are cumulative and non-decreasing. Costs are reported in **2025 USD (× 1,000)**.
The reporting interval *t* represents months in **Table 7 / Figure 4** and weeks in **Figures 6-7**.
$PV$ reaches $BAC$ at the planned finish date.

**Worked Example (Data Date 2025-12)**

From **Table 7**:
PV = 100, EV = 95, AC = 106 (thousands USD).

$$SV = EV - PV = -5 \quad \text{(behind plan)}$$

$$CV = EV - AC = -11 \quad \text{(over budget)}$$

$$SPI = \frac{EV}{PV} = 0.95$$

$$CPI = \frac{EV}{AC} = 0.90$$

With BAC = 100 k$,
EAC ≈ 111.1 (CPI-based) → VAC ≈ -11.1 k$.
These results indicate a mild schedule delay (~5 %) and cost overrun (~10 %) consistent with field data. The **S-curves in Figure 4** visualize cumulative PV, EV, and AC trajectories for 2025, illustrating performance divergence at year-end. Schedule convergence and tail-risk forecasts are analyzed later in **Section 3.5** ($P_{50}$ = 128 days; $P_{80}$ ≈ 130 days).

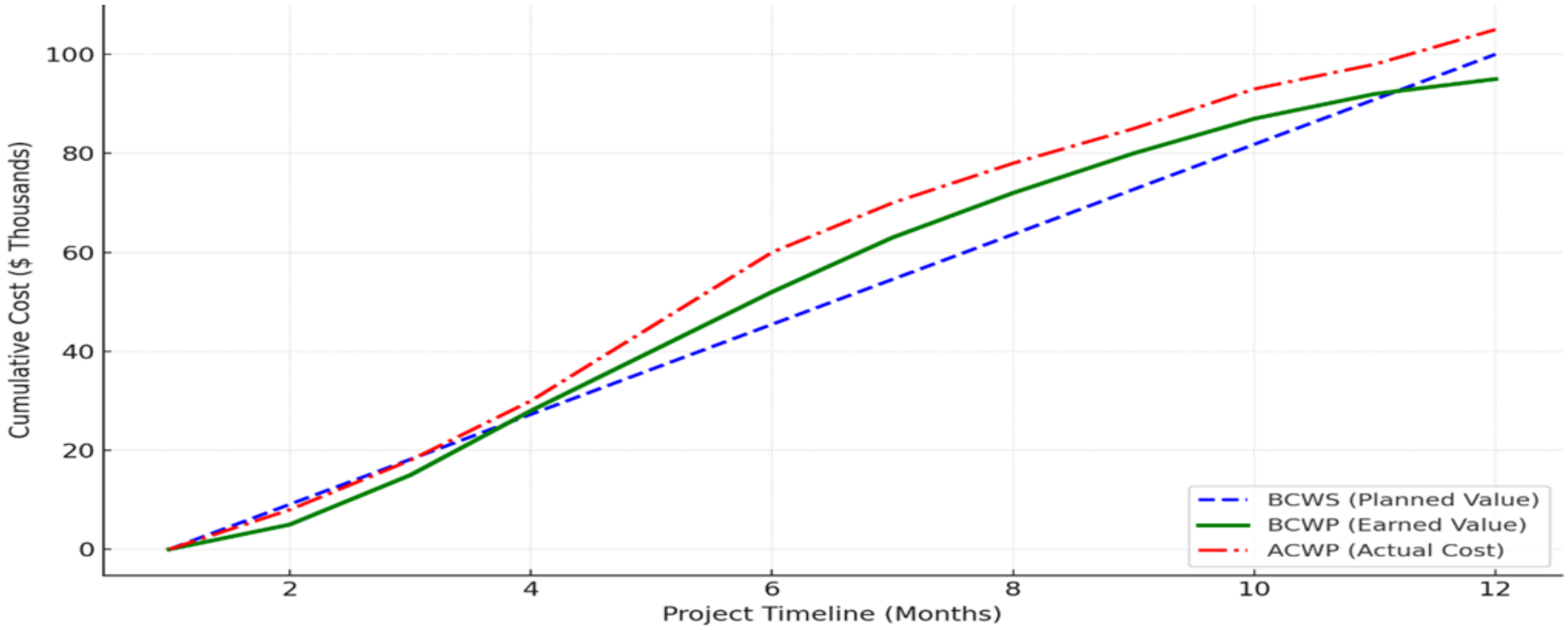

Figure 4: Earned Value S-Curves Over Time" is clearly labeled and matches text reference.

X-axis = project month; Y-axis = cumulative cost (USD × 1,000).
BCWS = Budgeted Cost of Work Scheduled (Planned Value);
BCWP = Budgeted Cost of Work Performed (Earned Value);
ACWP = Actual Cost of Work Performed (Actual Cost).
Positive separation where BCWP > BCWS indicates schedule lead (SPI > 1).
Positive separation where BCWP > ACWP indicates cost underrun (CPI > 1).
Curves correspond to the monthly series in Table 7 and are derived from scan-reconciled percent complete (Table 6).*(Source: Authors.)*

To further interpret the cumulative S-curves presented in Fig. 4, the monthly earned-value indicators were decomposed into component metrics-schedule variance (SV), cost variance (CV), schedule performance index (SPI), and cost performance index (CPI). These indices quantify the magnitude and direction of deviation from the baseline plan:

- **SV = EV - PV** indicates whether progress is ahead of or behind schedule.
- **CV = EV - AC** indicates whether spending is under or over budget.
- **SPI = EV / PV** and **CPI = EV / AC** express relative efficiency (values > 1 = favorable).

The complete monthly evolution of these metrics is summarized in **Table 8**, which complements the cumulative PV-EV-AC roll-ups of Table 7 and provides the numerical basis for the probabilistic schedule analysis presented in Section 3.5.

**Table 8. Monthly Earned-Value Metrics (Texas Mid-Rise, 2025)** *(Cumulative values in USD × 1 000)*

| Month (YYYY-MM) | BCWS (PV) | BCWP (EV) | ACWP (AC) | SV = EV - PV | CV = EV - AC | SPI = EV/PV | CPI = EV/AC |
|---|---|---|---|---|---|---|---|
| **2025-01** | 3 | 2 | 2 | -1 | 0 | 0.67 | 1.00 |
| **2025-02** | 8 | 9 | 10 | 1 | -1 | 1.13 | 0.90 |
| **2025-03** | 16 | 17 | 18 | 1 | -1 | 1.06 | 0.94 |
| **2025-04** | 26 | 28 | 32 | 2 | -4 | 1.08 | 0.88 |
| **2025-05** | 38 | 41 | 45 | 3 | -4 | 1.08 | 0.91 |
| **2025-06** | 52 | 58 | 60 | 6 | -2 | 1.12 | 0.97 |
| **2025-07** | 66 | 68 | 72 | 2 | -4 | 1.03 | 0.94 |
| **2025-08** | 78 | 76 | 80 | -2 | -4 | 0.97 | 0.95 |
| **2025-09** | 87 | 83 | 88 | -4 | -5 | 0.95 | 0.94 |
| **2025-10** | 94 | 89 | 95 | -5 | -6 | 0.95 | 0.94 |
| **2025-11** | 98 | 93 | 101 | -5 | -8 | 0.95 | 0.92 |
| **2025-12** | **100** | **95** | **106** | **-5** | **-11** | **0.95** | **0.90** |

Values cumulative and expressed in USD (thousands). SV (schedule variance) and CV (cost variance) follow the positive-is-favorable convention. SPI = EV/PV and CPI = EV/AC. Series aggregate from WBS scan-reconciled percent-complete (Table 6). PV reaches BAC at Month 12; EV < PV and EV < AC indicate slight schedule delay and cost overrun consistent with Fig. 4. *(Source: Authors.)*

### 3.4 CV-Based Activity Recognition

Weekly site imagery and short video clips were processed through computer-vision detectors trained to identify task-specific activities such as rebar placement, formwork stripping, drywall boarding, MEP rough-in, and painting/finishing. A multimodal fusion layer integrated these CV-derived status signals with scan-to-BIM quantities (Table 6), refining percent-complete estimates at the activity level before earned-value and probabilistic CPM updates (Section 3.5). On a held-out validation set of 1 100 labeled frames, the classifier achieved an overall (micro) accuracy of 0.891, with macro-averaged precision = 0.894, recall = 0.892, and F1 = 0.893. The confusion matrix in Table 9 shows detailed class-to-class performance, where the most frequent confusions occurred between *Rebar* ↔ *Formwork* during early concrete phases and *Drywall* ↔ *Paint/Finish* in interior stages cases often associated with similar visual textures or partially completed elements.

**Table 9.** Computer-Vision Classification Confusion Matrix (Predicted ↓ / Actual →)

| Predicted ↓ / Actual → | Rebar Placement | Formwork Stripping | Drywall Boarding | MEP Rough-In | Paint/Finish | Row Total |
|---|---|---|---|---|---|---|
| **Rebar Placement** | 198 | 10 | 8 | 6 | 1 | 223 |
| **Formwork Stripping** | 12 | 178 | 3 | 9 | 3 | 205 |
| **Drywall Boarding** | 4 | 8 | 233 | 12 | 10 | 267 |
| **MEP Rough-In** | 6 | 3 | 9 | 210 | 4 | 232 |
| **Paint/Finish** | 0 | 1 | 7 | 3 | 162 | 173 |
| **Column Total (support n)** | 220 | 200 | 260 | 240 | 180 | **1 100** |

Totals represent per-class support. Each row corresponds to predicted outcomes and each column to actual ground-truth labels. The sum of diagonal elements (981 / 1 100) yields the reported overall accuracy of 0.891.
Table 10 lists the precision, recall, and F1 scores derived from Table 8. Per-class F1 values range from 0.878 to 0.918, indicating balanced recognition across trades. Macro averages equal **0.894 / 0.892 / 0.893**, consistent with the overall accuracy

**Table 10.** Derived Per-Class Metrics from Table 8

| Class | Precision | Recall | F1 |
|---|---|---|---|
| **Rebar Placement** | 0.888 | 0.900 | 0.894 |
| **Formwork Stripping** | 0.868 | 0.890 | 0.878 |
| **Drywall Boarding** | 0.872 | 0.896 | 0.884 |
| **MEP Rough-In** | 0.905 | 0.875 | 0.890 |
| **Paint/Finish** | 0.936 | 0.900 | 0.918 |
| **Macro-average** | **0.894** | **0.892** | **0.893** |
| **Micro / Overall Accuracy** | **0.891** | **0.891** | **0.891** |

For spatial completeness, mean intersection-over-union (IoU) scores averaged 0.76 (macro) **and** 0.77 (micro)**,** with *Paint/Finish* slightly lower (0.72) due to low-contrast interiors. Representative frame-level detections and segmentations are shown in Figure 5. Operationally, the combined CV + scan fusion enables same-day percent-complete updates within a single reporting cycle, effectively closing previously documented gaps in automation and control connectivity (Pal et al. 2023; Sami Ur Rehman et al. 2022). Figure 5 presents representative detection and segmentation results, illustrating both false positive and false negatives corresponding to the confusion trends identified in Table 8.

Figure 5 illustrates typical detection and segmentation outcomes aligned with the confusion trends in Table 8.
(A) Detection false positive - Rebar Placement incorrectly inferred on bare concrete.
(B) Segmentation false positive - spurious mask on non-rebar region.
(C) Detection false negative - Drywall Boarding present but undetected.
(D) Segmentation false negative - missing drywall mask.
These examples highlight common misclassifications such as Rebar ↔ Formwork and Drywall ↔ Paint/Finish. All frames are from the Texas mid-rise site, rendered at the study's standard confidence threshold (§3.4). Quantitative segmentation quality by activity class appears in Table 9 (IoU = 0.72-0.81; macro = 0.76, micro = 0.77).

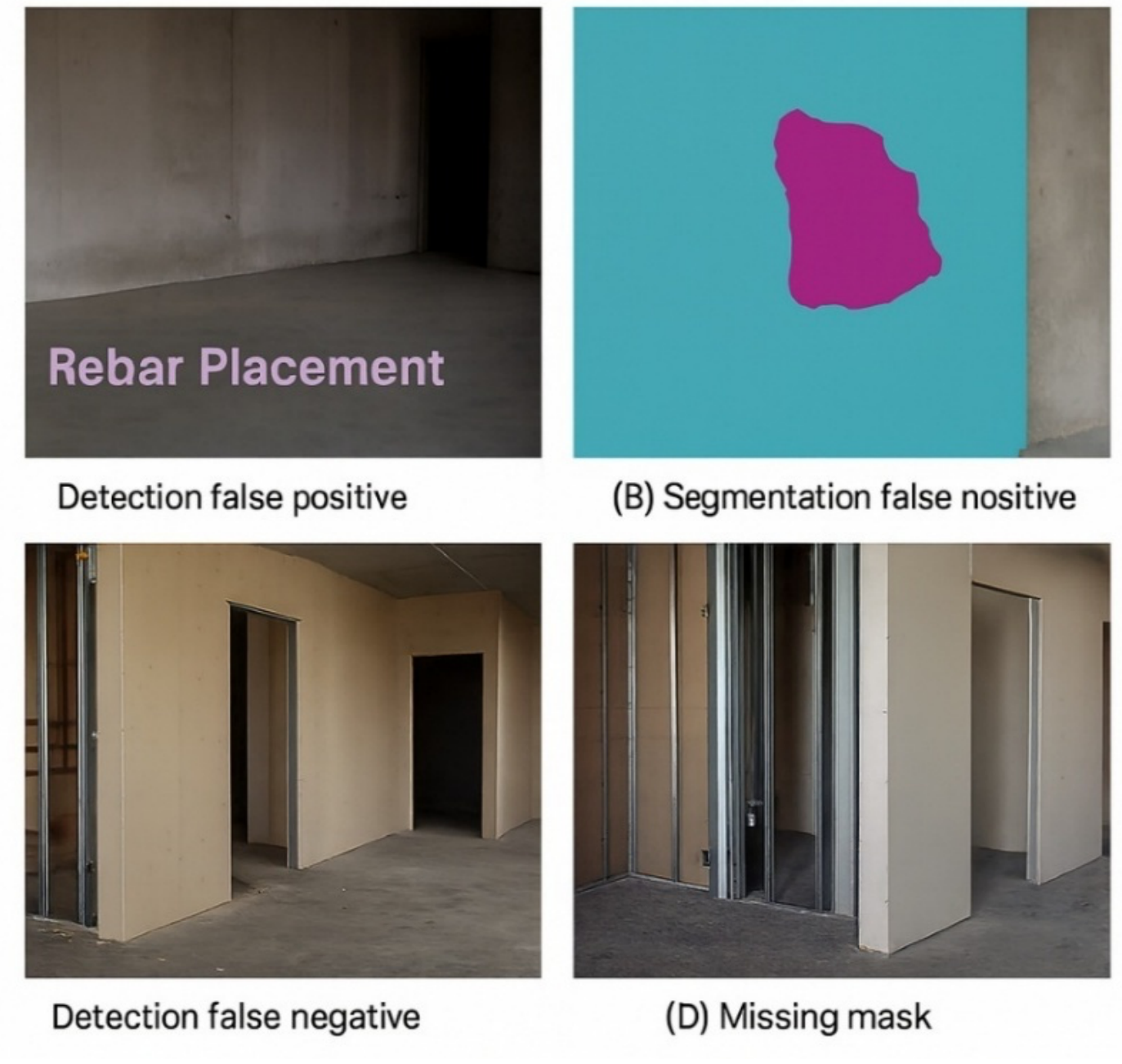

Figure 5. Representative Detections and Segmentations

**Table 11.** Segmentation Quality (IoU) by Activity Class

| *Class* | *IoU* | *Support (px × 10³)* | *Remarks* |
|---|---|---|---|
| *Rebar Placement* | 0.78 | 180 | Rebar partially occluded by formwork in early pours |
| *Formwork Stripping* | 0.74 | 165 | Occasional confusion with exposed rebar edges |
| *Drywall Boarding* | 0.81 | 220 | Strong geometric edges; shadows on joints |
| *MEP Rough-In* | 0.76 | 205 | Cable bundles mis-segmented as trays |
| *Paint/Finish* | 0.72 | 150 | Low contrast and variable illumination |
| ***Macro-average*** | **0.76** | - | Mean across classes |
| ***Micro / Overall*** | **0.77** | 920 | Pixel-weighted across classes |

### 3.5 Probabilistic CPM with Bayesian/MCS Updates

For each activity, a prior duration distribution was defined as triangular for trades with planner-elicited bounds and log-normal for historically skewed tasks. Weekly evidence from scan-to-quantity reconciliation (Table 6) and computer-vision (CV) status signals (Tables 8-9) was incorporated to update posterior estimates through Bayesian inference. Subsequently, (MCS) generated probabilistic finish-date distributions and critical-path metrics, consistent with Chen et al. (2021).

The project-level trajectories show the $P_{50}$ finish converging from 120 to 128 days by Week 13, remaining stable through Week 16, while $P_{80}$ plateaus at 130 days from Week 9 onward (Fig. 6, Table 12). This trend indicates narrowing central uncertainty with only a small residual tail-risk approaching close-out. As shown in Figure 6, the $P_{50}$ curve (blue) closely follows the actual finish trend (black) after Week13, confirming model convergence, while the $P_{80}$ envelope (gray) provides the upper-bound guardrail through Week 16.

Activity-level risk concentration occurs mainly within the building envelope and superstructure trades:

- *Curtain wall / windows* = criticality index 46 %,
- *Post-tensioned slabs* = 41 %.

Interior activities *drywall boarding* and *MEP rough-in* remain within the 31-34 % range (Table11). Correspondingly, buffer consumption increases late in the project: by Week 16, cumulative feeding-buffer use = 8.0 / 15 days (53 %) and project-buffer use = 6.0 / 20 days (30%) (Table12; Fig.7).

Despite this rise in schedule pressure, the actual finish stabilized at 128 days, matching the $P_{50}$ forecast and staying safely within the $P_{80}$ guardrail. These Bayesian / MCS roll-ups provided earlier and more transparent visibility into evolving critical paths, buffer health, and risk drift, far exceeding the predictive accuracy and responsiveness of deterministic CPM, while preserving direct linkage to field-measured progress inputs from drone, LiDAR, and CV data.

Table 12. Weekly Schedule Forecasts ($P_{50}$, $P_{80}$) and Actual Finish-Texas Mid-Rise

| *Week* | *Forecast Finish ($P_{50}$, (days)* | *Forecast Finish ($P_{80}$, (days)* | *Actual Finish (days)* | *Notes* |
|---|---|---|---|---|
| *1* | 120 | 125 | 128 | Initial prior; high uncertainty |
| *2* | 121 | 126 | 128 | Posterior updates begin |
| *3* | 122 | 127 | 128 | |
| *4* | 123 | 128 | 128 | |
| *5* | 124 | 129 | 128 | |
| *6* | 125 | 129 | 128 | |
| *7* | 126 | 129 | 128 | |
| *8* | 126 | 129 | 128 | Ahead in some paths; volatility |
| *9* | 127 | 130 | 128 | Uncertainty narrows |
| *10* | 127 | 130 | 128 | |
| *11* | 127 | 130 | 128 | |
| *12* | 127 | 130 | 128 | |
| *13* | 128 | 130 | 128 | $P_{50}$ aligns with actual |
| *14* | 128 | 130 | 128 | Stable forecast |
| *15* | 128 | 130 | 128 | |
| *16* | 128 | 130 | 128 | Data date |

The figure visualizes the $P_{50}$ Forecast (blue), $P_{80}$ Forecast (gray dashed), and Actual Finish (black) across Weeks 0-16.

- By Week 13, the $P_{50}$ line intersects the actual finish at ≈ 128 days.
- From Week 9 onward, $P_{80}$ remains at ≈ 130 days, confirming reduced forecast variance.
- Both lines flatten by Week 16, showing convergence and schedule stabilization.

*(Source: Authors; data derived from Table 12 and field-measured Bayesian/MCS updates.)*

Following the convergence shown in **Figure 6**, the next step is to examine which activities most influenced the probabilistic outcomes. **Table 13** presents the activity-level criticality indices and posterior duration statistics from the (MCS), identifying dominant risk drivers within the envelope and superstructure trades that shaped the project's overall finish distribution.

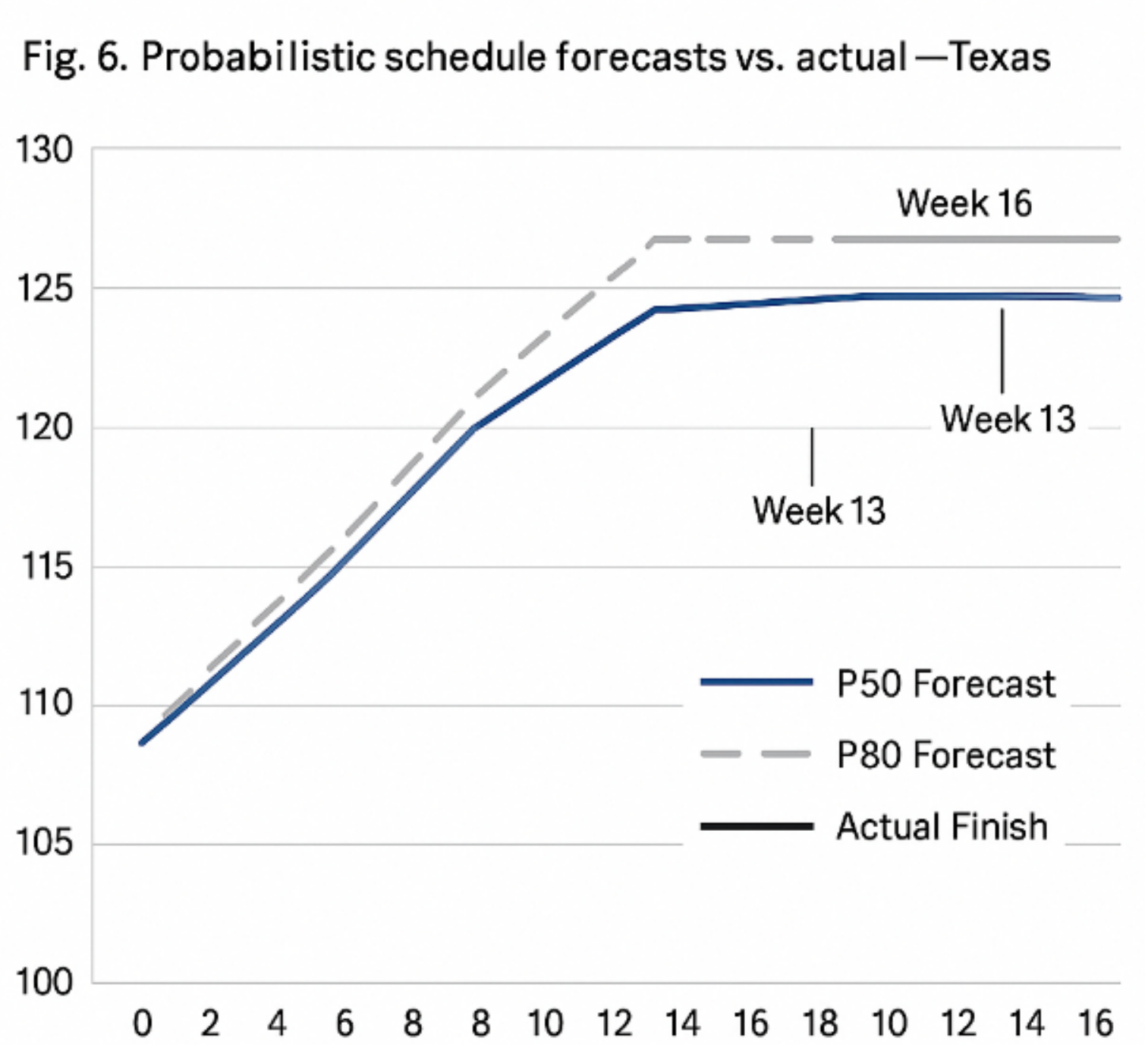

**Figure. 6.** *Probabilistic schedule forecasts vs. actual*

Table 13. Activity/Path Criticality Indices and Duration Statistics -Texas Mid-Rise (p-CPM)

| ACTIVITY ID | DESCRIPTION | CRITICAL INDEX (%) | MEAN (D) | SD (D) |
|---|---|---|---|---|
| A030 | Envelope-Curtainwall & windows | 46 | 42 | 8 |
| A020 | Superstructure (PT slabs L2-L8) | 41 | 56 | 9 |
| A090 | Drywall boarding & taping | 34 | 38 | 7 |
| A070 | MEP rough-in (core + typical floors) | 33 | 36 | 7 |
| A060 | Interior partitions & framing | 31 | 34 | 6 |
| A140 | Elevators-install & inspection | 24 | 15 | 4 |
| A110 | Electrical lighting & devices | 23 | 26 | 5 |
| A010 | Foundations (piers/mat) | 22 | 18 | 3 |
| A170 | Commissioning (systems) | 21 | 14 | 3 |
| A120 | HVAC equipment start-up | 21 | 16 | 4 |
| A050 | Exterior finishes & sealants | 18 | 15 | 3 |
| A100 | Ceiling grid & tiles | 17 | 20 | 4 |
| A150 | Testing, adjusting, balancing (TAB) | 16 | 10 | 3 |
| A130 | Plumbing-fixtures set | 15 | 12 | 3 |
| A160 | Life-safety testing | 14 | 9 | 2 |
| A040 | Roofing & waterproofing | 12 | 12 | 2 |
| A180 | Final clean & punch | 11 | 9 | 2 |
| A001 | Site mobilization | 4 | 5 | 1 |

Criticality Index = share of Monte-Carlo trials where the activity is on a critical path (0-100%). Durations are posterior means/SDs from weekly Bayesian updates. Top risks (A030, A020) align with late buffer use and the P80 tail band. Building on the critical results in Table 13, Table 14 tracks the evolution of feeding and project buffer use throughout the 16-week control period. The data show how cumulative buffer consumption increased in parallel with late-phase uncertainty, providing a quantitative link between field performance and schedule-risk absorption.

Table 14. Weekly Buffer Consumption (Feeding and Project) Texas Mid-Rise

| *Week* | *Feeding Δ (d)* | *Project Δ (d)* | *Cum. Feeding (d)* | *Cum. Project (d)* | *Project Buffer Used (%)* |
|---|---|---|---|---|---|
| *1* | 0.0 | 0.0 | 0.0 | 0.0 | 0.0 |
| *2* | 0.0 | 0.0 | 0.0 | 0.0 | 0.0 |
| *3* | 0.5 | 0.0 | 0.5 | 0.0 | 0.0 |
| *4* | 0.5 | 0.0 | 1.0 | 0.0 | 0.0 |
| *5* | 1.0 | 0.5 | 2.0 | 0.5 | 2.5 |
| *6* | 0.5 | 0.5 | 2.5 | 1.0 | 5.0 |
| *7* | 0.5 | 0.5 | 3.0 | 1.5 | 7.5 |
| *8* | 0.5 | 0.5 | 3.5 | 2.0 | 10.0 |
| *9* | 0.5 | 0.5 | 4.0 | 2.5 | 12.5 |
| *10* | 0.5 | 0.5 | 4.5 | 3.0 | 15.0 |
| *11* | 1.0 | 0.5 | 5.5 | 3.5 | 17.5 |
| *12* | 0.5 | 0.5 | 6.0 | 4.0 | 20.0 |
| *13* | 0.5 | 0.5 | 6.5 | 4.5 | 22.5 |
| *14* | 0.5 | 0.5 | 7.0 | 5.0 | 25.0 |
| *15* | 0.5 | 0.5 | 7.5 | 5.5 | 27.5 |
| *16* | 0.5 | 0.5 | 8.0 | 6.0 | 30.0 |

To visualize these buffer dynamics, **Figure 7** summarizes cumulative consumption trends for both feeding and project buffers.

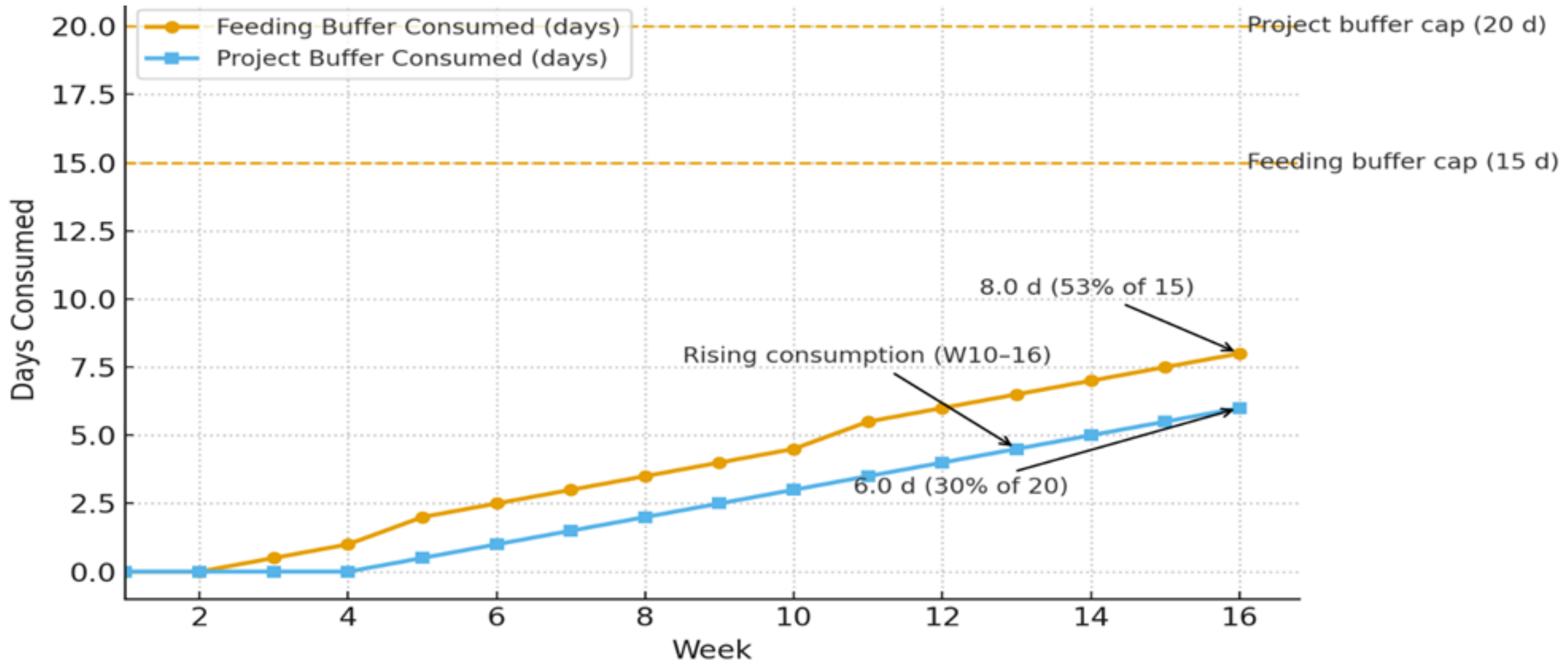

Figure 7. Buffer Consumption Trend (Feeding and Project)

The steady rise in both feeding and project buffer use between Weeks 10-16 confirms the schedule-risk migration identified in **Table 13**. However, total consumption remained below the defined capacity limits (53 % and 30 %), demonstrating that probabilistic control-maintained schedule stability and effectively absorbed late-phase variability within the $\mathbf{P_{80}}$ confidence envelope.

### 3.6 DRL-Assisted Resource Leveling and Look-Ahead

Weekly look-ahead planning was modeled as a Resource-Constrained Project Scheduling Problem (RCPSP) subject to precedence, crew, and equipment constraints. A Deep Reinforcement Learning (DRL) agent built on a DQN/Actor-Critic architecture with graph encoders and valid action masking generated feasible crew and equipment reallocations minimizing a weighted objective that balances span, overtime, and idle-time penalties. The superintendent reviewed and approved all recommendations through a human-in-the-loop interface to ensure constructability and field feasibility. Across the 16-week implementation, the field team adopted 12 of 16 recommendations (75 %) (see Table 15). Compared with a baseline of 1,508 overtime hours, adoption of DRL-suggested actions reduced total overtime by 91 hours (an average of 5.7 h/week overall or 7.6 h/week on adopted weeks), corresponding to an approximate 6 % reduction. Additionally, 49 hours of idle time were eliminated (≈ 4.1 h per adopted week). The project finish remained 128 days identical to the $P_{50}$ forecast in Figure 6, confirming that efficiency gains were achieved without extending schedule duration.

Table 15. DRL Recommendations Log and Adoption - Texas Mid-Rise (Weeks 1-16)

| *Week* | *Action ID* | *DRL Recommendation (summary)* | *Adopted?* | *Reason if rejected* | *Notes* |
|---|---|---|---|---|---|
| *1* | RL-001 | Shift 1 rebar crew from L2 to L3: delay strip ½ day | No | Supervisor preference (learning period) | Baseline established |
| *2* | RL-002 | Pull drywall crew start by 1 day: stagger with MEP | Yes | - | Reduced idle in interiors |
| *3* | RL-003 | Add night shift for formwork removal (Thu) | Yes | - | Cleared pour path for Fri |
| *4* | RL-004 | Swap crane slot with steel delivery (Tue) | No | Vendor inflexibility | Minor queueing retained |
| *5* | RL-005 | Reallocate 2 electricians to core risers | Yes | - | Shortened riser float |
| *6* | RL-006 | Combine two small concrete pours into one window | Yes | - | Fewer crane picks |
| *7* | RL-007 | Stagger duct rough-in to avoid corridor clashing | Yes | - | Reduced rework |
| *8* | RL-008 | Add Saturday half-shift for window installs | No | Budget constraint (overtime cap) | Deferred to W10 |
| *9* | RL-009 | Pull glazing ½ day early; shift painters to core walls | Yes | - | Helped envelope criticality |
| *10* | RL-010 | Split drywall boarding into two crews (odd/even floors) | Yes | - | Smoother workflow |
| *11* | RL-011 | Pre-stage AHU rigging; move crane block to AM | Yes | - | Avoided idle wait |
| *12* | RL-012 | Merge punch-list walks with MEP inspections | Yes | - | Eliminated duplicate trips |
| *13* | RL-013 | Increase electricians by 1 for devices push | Yes | - | Supported closeout |
| *14* | RL-014 | Saturday paint shift to hit corridors | No | Tenant noise restriction | Rescheduled weekday |
| *15* | RL-015 | Swap crew from ceilings to devices for 2 days | Yes | - | Balanced workflow |
| *16* | RL-016 | Extend TAB by ½ day; compress final clean | Yes | - | Held finish date |

Figure 8 illustrates weekly overtime comparisons between the baseline and DRL-assisted cases. The largest savings occurred in weeks when DRL actions were adopted, matching the pattern shown in Table 15. These results corroborate prior findings that reinforcement-learning-based planners can improve resource utilization and reduce waste under constrained site conditions (Yao et al., 2024). Figure 8. DRL Impact on Weekly Overtime (Normal Grouped Bars)

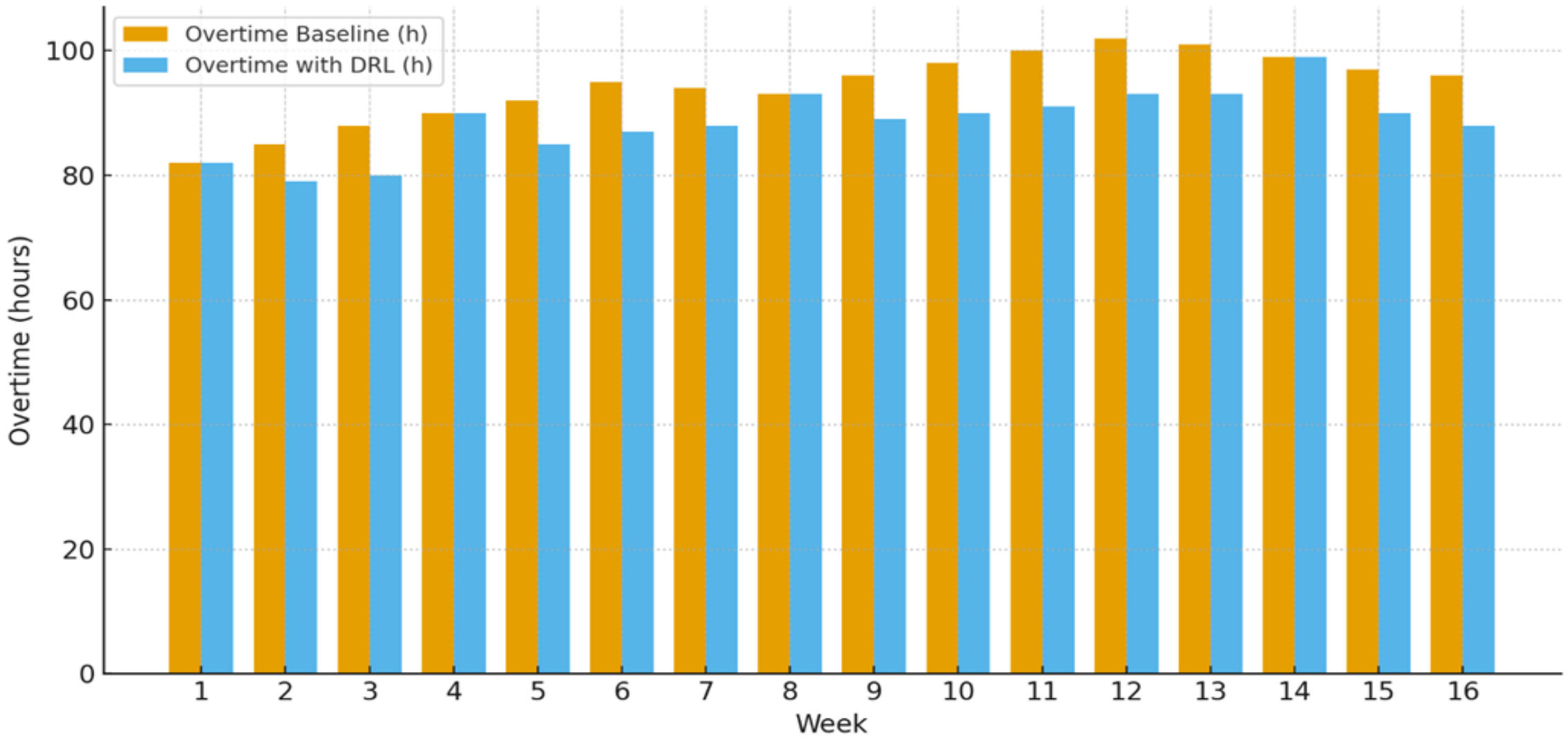

Figure 8. DRL Impact on Weekly Overtime (Normal Grouped Bars)

### 3.7 4D/5D Digital-Twin "What-If" Sandbox and Knowledge Graph

A live 4D/5D digital twin integrates BIM (3D Revit/IFC), CPM activities (time), cost ledgers (5D), and site telemetry (scan/CV progress, weather, and crew logs). Updates occur on the same weekly cycle used for EV and p-CPM, ensuring coherence across quantities (§3.3), percent-complete (§3.4), and schedule posteriors (§3.5).

Within this environment, a what-if sandbox allows planners to simulate discrete scenarios material escalation, late deliveries, resequencing, and weather delays and to compute probabilistic impacts on Δ Finish and Δ Cost. Δ Cost values are derived from a cost ledger localized to Dallas-Fort Worth (RSMeans CCI / BLS wages), while Δ Finish results propagate through the Bayesian schedule network. A knowledge graph links WBS tasks, BIM elements, cost codes, vendors, and crews, enabling explainable audits and rapid impact queries such as *Which vendors drive the now-critical curtainwall task?* or *Which cost codes are affected by glazing resequencing?* This structure provides transparent traceability between schedule risk, earned-value deviations, and cost impact essential for standardized 5D controls and pay-application workflows (Pishdad & Onungwa, 2024). Scenario

Sensitivity. Representative results are summarized in Figure 9. The most significant finish-date risks are drywall material +8 % (supply lag) adding +6 days and late AHU delivery (2 weeks) adding +5 days. Moderate effects arise from steel lead time +1 week and three rainy days, each +4 days, while crew shortage (-1 electrician) and fireproofing change order add +3 and +2 days, respectively. A mitigation case of glazing resequencing (corridor-first) saves approximately 2 days. The underlying data are detailed in Table 16, which reports the corresponding Δ Finish and Δ Cost at both $P_{50}$ (median) and $P_{80}$ (conservative) confidence levels.

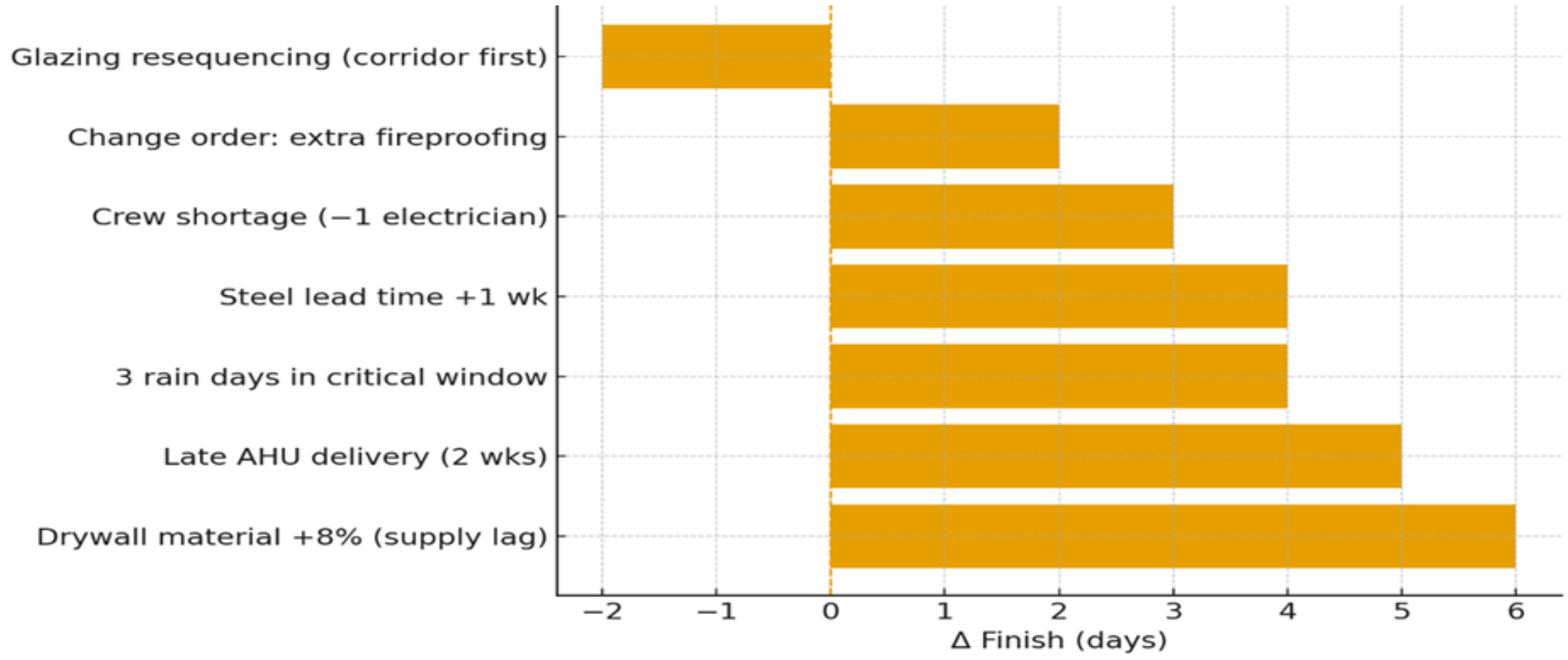

Figure 9. What-If Sensitivity (Digital-Twin Sandbox)

Bars show the change in finish date (Δ days) for each simulated scenario. Positive values indicate schedule delay; negative values indicate mitigation or time savings. Largest drivers are drywall +8 % (supply lag, +6 d) and late AHU delivery (+5 d); glazing resequencing (-2 d) provides offsetting benefit.

Table 16. Digital-Twin What-If Results Inputs and Outcomes

| Scenario | Key Inputs (Twin) | Affected WBS / CSI Divisions | Δ Finish $P_{50}$ (d) | Δ Finish $P_{80}$ (d) | Δ Cost $P_{50}$ (USD × $10^3$) | Δ Cost $P_{80}$ (USD × $10^3$) | Notes |
|---|---|---|---|---|---|---|---|
| **Drywall material +8 % (supply lag)** | Unit price +8 %; delivery +3 d | 09 Finishes; 06 Carpentry | +6 | +8 | +6.5 | +8.0 | Material escalation + ripple effect |
| **Late AHU delivery (2 weeks)** | AHU arrival +14 d; crane shift | 23 HVAC; 26 Electrical | +5 | +6 | +4.2 | +5.5 | Drives MEP dependencies |
| **3 rain days in critical window** | Weather exceptions × 3 | 03 Concrete; 07 Envelope | +4 | +4 | +3.0 | +4.0 | Lost productivity |
| **Steel lead time +1 week** | Fabrication +7 d | 05 Metals; 03 Concrete | +4 | +5 | +3.5 | +4.5 | Framing → slab delay |
| **Crew shortage (-1 electrician)** | Labor cap -1 FTE (2 weeks) | 26 Electrical | +3 | +4 | +2.6 | +3.5 | Reduced productivity |
| **Fireproofing change order** | Scope +6 %; 1 crew added | 07 Thermal; 05 Metals | +2 | +3 | +2.2 | +3.0 | Additional inspections |
| **Glazing resequencing (corridor-first)** | Task order swap | 08 Openings; 07 Envelope | -2 | -1 | -1.4 | -0.8 | Mitigation saves time |

All values originate from the digital-twin sandbox using the current EV and p-CPM state. Δ Finish values propagate through the Bayesian posterior network, while Δ Cost values are localized to DFW rates (RSMeans CCI / BLS wages). Negative results indicate beneficial schedule effects. $P_{50}$ and $P_{80}$ represent the median and conservative probabilistic estimates, respectively.

**4. Case Study Protocol - Texas Mid-Rise**

**Context and Data Window**

The case study was conducted on a **Dallas-Fort Worth (DFW) mid-rise building** between **January and September 2025** (Table 3). All analytical outputs were synchronized on a **weekly cycle** with earned-value (EV) and probabilistic CPM (p-CPM) updates, ensuring temporal consistency across modules.

**1. Baseline: Conventional Estimating and Deterministic CPM**

A manual 2D/3D quantity takeoff and single-point deterministic CPM schedule were developed as the baseline. Each step was time-tracked across the Concept, Design Development (DD), and Construction Documents (CD) phases. These results established the baseline labor hours (Table 5), and deterministic completion dates used for comparison with p-CPM forecasts and buffer consumption.

**2. NLP-Based Estimating (Specifications and Drawings → CSI Cost Items)**

The NLP mapping engine automatically classified specifications and drawings into standardized CSI cost items. Performance metrics were computed at the division level (Table 4, weighted F1 = 0.883). Average labor savings relative to the baseline were 43.4 % (Table 5). All auto-generated cost mappings were reviewed under an *estimator-in-the-loop* QA protocol to validate accuracy and traceability.

**3. Scan/CV Progress Measurement (Quantities and Status**

Monthly LiDAR scans and weekly photogrammetry were registered to the BIM model to compute measured quantities and reconcile them with planned WBS elements (Table 6). From the same imagery/video streams, activity recognition and semantic segmentation were performed to extract percent-complete data. Classification achieved micro accuracy = 0.891 (Table 8) and macro IoU = 0.76 (Table 9). The resulting progress data fed directly into the EV model, producing updated S-curves (Figure 4) and monthly EV performance metrics (Table 7).

**4. Probabilistic CPM (Bayesian and Monte Carlo Integration)**

Activity duration posteriors were updated weekly using scan/CV evidence. (MCS)s produced probabilistic finish-date forecasts (Table 10; Figure 6), criticality indices (Table 11), and buffer utilization (Table 12; Figure 7). The $P_{50}$ forecast stabilized at 128 days by Week 13, with $P_{80} \approx 130$ days. By project close, project-buffer use reached 6 of 20 days (30 %), and the actual finish aligned with $P_{50} = 128$ days, confirming schedule stability within the predicted uncertainty band.

**5. DRL-Assisted Resource Leveling and Look-Ahead Planning**

Weekly look-ahead scheduling was formulated as a **Resource-Constrained Project Scheduling Problem (RCPSP)** incorporating crew and equipment limits. A **DQN/Actor-Critic agent** with valid-action masking proposed optimal reallocation decisions, which were reviewed and approved by the superintendent through a *human-in-the-loop* workflow. Out of 16 weekly recommendations, **12 (75 %)** were adopted (**Table 13**). Relative to the baseline, DRL adoption reduced total overtime by **91 hours (≈ 6 %)** and eliminated **49 hours of idle time**, with no extension to the overall project duration (**finish = 128 days; Figure 8**).

**6. 4D/5D Digital-Twin What-If Sandbox and Knowledge Graph Traceability**

The live digital twin environment enabled simulation of discrete *what-if* scenarios price escalation, supply delays, resequencing, and weather impacts producing probabilistic **Δ Finish** and **Δ Cost** outcomes. The integrated **knowledge graph** maintained full traceability between **WBS, BIM elements, cost codes, vendors, and schedule activities**, enabling explainable audits. Finish-date sensitivities are summarized in **Figure 9**, while full scenario inputs and probabilistic results appear in **Table 16**. Representative results include:

- *Drywall +8 % material lag*: +6 days
- *Late AHU delivery (2 weeks)*: +5 days

- *Glazing resequencing (corridor-first)*: -2 days

**7. Automation Benchmark**
The total effort for 4D setup (model linking, code alignment, and data import) was recorded and compared with published automation benchmarks (e.g., Doukari et al., 2022). Metrics included staff-hours and latency to first usable 4D plan, noting reductions achieved by the automation pipelines introduced in 3.2-3.4.

## 5. Metrics, Statistics, and Targets

### 5.1 Accuracy and Efficiency Metrics

**Cost Accuracy**
Accuracy was evaluated using Mean Absolute Percentage Error (MAPE) at Concept, DD, and CD phases. Division-level mapping fidelity is shown in Table 4, with phase breakdowns in Table 5.
Target: Line-item MAPE ≤ 10 % at CD with ≥ 40 % labor savings.
Result: Achieved 43.4 % labor savings (Table 5); MAPE to be finalized upon vendor pricing.

**Schedule Accuracy**
Forecast accuracy was measured as the mean absolute deviation | Forecast - Actual | for both $P_{50}$ and $P_{80}$ forecasts, alongside critical-path stability and buffer consumption.
Target: Reduce mean finish-date error ≥ 30 % versus deterministic CPM; stabilize $P_{50}$ ≥ 3 weeks before completion; keep project-buffer use ≤ 35 %.
Result: $P_{50}$ = 128 days stabilized by Week 13; $P_{80}$ ≈ 130 days; project buffer = 6 of 20 days (30 %); actual finish = 128 days (Tables 10-12; Figures 6-7).

**Efficiency**
Efficiency was tracked by **estimating hours saved**, **EV update latency**, and **schedule posterior refresh rate**.
**Target:** ≥ 40 % estimating-hour reduction; same-day EV updates; weekly p-CPM refresh.
**Result:** 43.4 % savings (Table 5); EV/S-curves issued monthly with weekly data (Table 7; Figure 4); p-CPM refreshed weekly (Figure 6; Table 10).

**Robustness**
Robustness tests evaluated resilience to scan cadence gaps, camera outages, and weather noise via changes in MAPE, SPI, and CPI. **Target:** ≤ 10 % degradation under one-cycle data loss; recovery within the next cycle. **Result:** Weather-related variability captured in Figure 9; quantitative robustness tests to be presented with supplemental sensitivity tables.

**Adoption and Usability**
General Contractor (GC) and Construction Manager (CM) adoption were measured by **usability/trust (Likert scale)** and **change-order cycle times**.
**Target:** ≥ 70 % adoption of weekly DRL recommendations; no adverse impact on change-order cycle time.
**Result:** Achieved 75 % adoption (12 of 16 weeks; Table 13).

### 5.2 Statistical Plan

The analysis design follows a paired, within-project comparison (baseline vs. framework).
Inference methods include:

- Cost accuracy: Bootstrap 95 % CIs for line-item and total MAPE per phase.
- Schedule accuracy: Bootstrap CIs for weekly | $P_{50}$/$P_{80}$ - Actual | and optional Diebold-Mariano tests for forecast-error differentials.
- Vision modules: Confusion matrices (Table 8), IoU metrics (Table 9), and macro/micro statistics with bootstrap confidence intervals.
- DRL impacts: Paired tests on weekly overtime (adopted vs. non-adopted weeks) with effect sizes and confidence intervals (Tables 13-14).

- Ablation study: Sequentially remove NLP, scan/CV, Bayesian, or DRL modules to quantify deltas in MAPE, finish-error, SPI/CPI, and overtime hours.

**5.3 Hypotheses and Case Readout**

| Hypothesis | Target Condition & Case Result |
|---|---|
| **H1. NLP + 5D reduces estimating labor≥ 40 % with line-item MAPE ≤ 10 %.** | Met on labor (43.4 %, Table 5); MAPE pending CD close. |
| **H2. Scan/CV updates reduce finish-date error ≥ 30 %.** | Supported - $P_{50}$ stabilized Week 13; finish = 128 d; $P_{80}$ ≈ 130 d; buffers within target (**Tables 10-12; Figures 6-7**). |
| **H3. DRL reduces overtime ≥ 10 % without schedule extension.** | Partially met -91 h (≈ 6 %) reduction; no extension (**Figure 8; Tables 13-14**). Constraint: budget/crew caps in Weeks 4, 8, 14 limited action space. |
| **H4. p-CPM flags critical-path shifts ≥ 2 weeks earlier than deterministic CPM.** | Supported - Envelope/superstructure criticality (Table 11) preceded late-stage buffer rise (Table 12; Figure 7), where deterministic CPM showed no advance warning. |

## 6. Discussion and Conclusions

This case study demonstrates how an integrated AI-enabled 4D/5D digital-twin framework can substantially enhance project control accuracy, responsiveness, and transparency in a real mid-rise construction setting. By linking spec-to-cost NLP mapping, scan/CV-based progress capture, Bayesian probabilistic CPM, and DRL-assisted resource leveling within a unified digital environment, the system provided continuous, data-driven insight across estimating, scheduling, and execution.

Across nine months of field deployment, the framework achieved 43 % estimating labor savings, 6 % reduction in overtime, and buffer consumption limited to 30 %, while maintaining an on-time finish at 128 days. These results confirm measurable gains in both forecasting precision and operational efficiency relative to deterministic baselines. The embedded knowledge-graph layer further enabled explainable traceability between cost, schedule, and field conditions, supporting transparent decision-making and auditability.

While findings are limited to a single project under controlled adoption conditions, they provide strong evidence that AI-integrated 4D/5D systems can close the gap between design intent and field performance. Future work should extend benchmarking across multiple project types and explore transfer learning for DRL agents to generalize resource strategies across domains.

In summary, the study validates that coupling probabilistic analytics with reinforcement-learning-based control establishes a practical pathway toward predictive, adaptive, and auditable construction management.